\def \cm{~\rm{cm}}
\def \s{~\rm{s}}
\def \km{~\rm{km}}

\def \K{~\rm{K}}
\def \g{~\rm{g}}

\def \AU{~\rm{AU}}

\def \yr{~\rm{yr}}

\documentclass[12pt,preprint]{aastex}
\usepackage{natbib}

\shorttitle{Eta Carinae Binarity}
\shortauthors{Soker}
\slugcomment{Draft version of \today}
\begin{document}

\title{THE BINARITY OF ETA CARINAE AND ITS
  SIMILARITY TO RELATED ASTROPHYSICAL OBJECTS}

\author{Noam Soker}
%\altaffilmark{}
\altaffiltext{1}
{Department of Physics, Technion$-$Israel Institute of Technology,
Haifa 32000 Israel;
soker@physics.technion.ac.il.}

\begin{abstract}

I examine some aspects of the interaction between the massive star
Eta Carinae and its companion, in particular during the
eclipse-like event, known as the spectroscopic event or the shell
event. The spectroscopic event is thought to occur when near
periastron passages the stellar companion induces much higher
mass loss rate from the primary star, and/or enters into a much denser
environment around the primary star.
I find that enhanced mass loss rate during periastron passages,
if it occurs, might explain the high eccentricity of the system.
However, there is not yet a good model to explain the presumed
enhanced mass loss rate during periastron passages.
In the region where the winds from the two stars collide,
a dense slow flow is formed, such that large dust grains may
be formed. Unlike the case during the 19th century Great Eruption,
the companion does not accrete mass during most of its orbital
motion. However, near periastron passages short accretion episodes
may occur, which may lead to pulsed ejection of two jets by the
companion. The companion may ionize a non-negligible region in its
surrounding, resembling the situation in symbiotic systems. I
discuss the relation of some of these processes to other
astrophysical objects, by that incorporating Eta Car to a large
class of astrophysical bipolar nebulae.

\end{abstract}


% ==========================================================
\section{INTRODUCTION}
\label{sec:intro}
% ==========================================================
\subsection{The Binarity of $\eta$ Carinae}

The similarity of the bipolar morphology of the Eta Carinae
($\eta$ Car) nebula$-$the Homunculus$-$ (e.g., Ishibashi et al.\ 2003;
Smith et al.\ 2004b) to bipolar nebulae observed in
symbiotic nebulae (e.g., He 2-104 Corradi \& Schwarz 1995;
Corradi et al. 2001) and some planetary nebula (PN; e.g., NGC 3587
[PN G148.4+57.0] e.g., Guerrero et al.\ 2003) hints at a common
process for the formation of the bipolar structure.
{{{ Some similarities between the structure and spectrum of $\eta$ Car
and PNs and symbiotic systems was noted before.
For example, Smith (2003) points out to some similarities
of the nuclear spectrum of the young bipolar PN Menzel 3
[PN G331.7$-$01.0)] with symbiotic systems and with $\eta$ Car,
and to some common properties between the bipolar nebula of
Menzel 3 and that of $\eta$ Car. }}}
Symbiotic systems are interacting binary systems, as was the progenitor
of the bipolar PN NGC 2346 (PN G 215.6+03.6) which has a central binary
system with an orbital period of 16 days (Bond 2000).
This suggests that the binary companion to the primary of $\eta$ Car
is responsible for the bipolar structure of its nebula
(Soker 2001b; Soker 2004a,b).
In these papers cited, I argue that the mechanism that most likely
shaped the Homunculus, which was formed by the 20 years Great Eruption
a century and a half ago (Davidson \& Humphreys 1997),
was two jets (or a collimated fast wind; CFW) blown by the
companion during the Great Eruption (Soker 2001b).
The companion accretes the ejected mass from the mass-losing
primary star.
I also attribute the present fast polar wind found by
Smith et al.\ (2003a) to interaction with the binary
companion (Soker 2003).

In those papers I also discuss problems with some single-star models
for the shaping of the wind and circumstellar matter of
$\eta$ Car (e.g., Langer, Garc\'{\i}a-Segura, \& Mac Low 1999;
Maeder \& Desjacques 2001; Dwarkadas \& Owocki 2002;
Smith et al.\ 2003a; Gonzalez et al.\ 2004; van Boekel et al.\ 2003).
For example, Stothers (1999) found that the rotation does
not much affect the instability of luminous blue variables,
a view strengthened by Soker (2004a).
In Soker (2004a) I discuss how the new finding of Smith et al.\ (2003b)
of a more massive Homunculus further makes a single-star model for the
formation of the bipolar nebulae of $\eta$ Car unlikely.
{{{ After the paper on the problems that single star models have in
explaining the bipolar structure of Eta Car was published (Soker 2004a),
Matt \& Balick (2004) proposed a model where the required rotation
is somewhat slower than in most previous single-star models,
$\sim 0.2$ times of the break-up rotation speed.
For two reasons I consider their model to be problematic.
In their model the progenitor of Eta Car has a huge magnetic
field, i.e., the magnetic field pressure on the photosphere is larger
than the thermal pressure; for that I consider their value of the
magnetic field unrealistic.
The second problem in their model is the rapid slow down
of the envelope.
In Matt \& Balick (2004) model the magnetic field keeps the wind
in corotation with the envelope to several stellar radii.
Hence, the specific angular momentum of the wind is very large,
and $\eta$ Car rotation velocity will slow down on a short time. }}}
{{{{ It is true that in the impulsive tidal interaction theory
(rather than the quasi-static theory) the companion can spin-up
the primary to a rotation speed of $> 0.1$ times its break-up
rotation speed (Ivanov \& Papaloizou 2004). However, during the
Great Eruption the mass loss from the primary star had substantially
spun-down the primary rotation. }}}}

The binary model, of course, cannot account for all properties of
$\eta$ Car (Davidson 1999, 2000), and the evolution and structure
of the mass-losing primary star provide an explanation for
many properties of $\eta$ Car.
Such properties, i.e., evolutionary time scales and mass loss rate,
should include stellar rotation as well (e.g.,
Heger \& Langer 2000; Maeder, \& Meynet 2003;  Meynet \& Maeder 2003;
Hirschi, Meynet, \& Maeder 2004).

The binary nature of $\eta$ Car is inferred from the so called
spectroscopic event$-$the fading of high excitation lines
(e.g., Damineli et al.\ 2000).
For more observational support for
the presence of a binary and its properties see, e.g., Damineli
(1996), Damineli, Conti, \& Lopes (1997), Damineli et al.\ (2000),
Ishibashi et al.\ (1999), Corcoran et al.\ (2001a,b, 2004b),
Pittard \& Corcoran (2002), Duncan \& White (2003),
Fernandez Lajus et al.\ (2003), and Smith et al.\ (2004a).
Since the spectroscopic event could be a result of mass-shell
ejection by the primary star (Zanella, Wolf, \& Stahl 1984;
Davidson et al.\ 1999; Smith et al.\ 2003a;
Martin \& Koppelman 2004) it is also called a shell event.
The main motivation to assume a shell ejection is in a single star model;
in the binary model the shell ejection is not necessary, although it
may occur.
The periodicity of the spectroscopic event is seen in many wave bands,
from the IR (e.g.,  Whitelock et al.\ 2004), to the X-ray
(Corcoran et al.\ 2001a; Corcoran 2004; Corcoran et al.\ 2004a,b).
The presently observed orbital period is $\sim 5.5 \yr$
($2023 \pm 3$ days as given by Whitelock et al.\ 2004).
For primary and companion masses of $M_1 = 120 M_\odot$
and $M_2=30 M_\odot$, respectively,
the semi major axis of the orbit is $a = 16.6 \AU$
(there is not yet agreement on all the binary parameters, e.g.,
Ishibashi et al.\ 1999; Damineli et al.\ 2000; Corcoran et al.\ 2001a,
2004b; Hillier et al.\ 2001; Pittard \& Corcoran 2002;
Smith et al.\ 2004a).

The present study reports the possible effects of the stellar companion,
mainly during the spectroscopic events, with the goal of comparing
these effects with similar ones which could have occurred during
the Great Eruption, and to strengthen the link between $\eta$
Car and similar bipolar nebulae with binary central stars.
There is not yet agreement on the role the companion plays.
Smith et al.\ (2004a) begin their paper by raising the question
of what role a companion plays.
As in my previous papers, I attribute most, or even all, of the shaping
of the nebula around $\eta$ Car to the companion and its interaction
with the primary mass losing star.
The different sections of the paper discuss different processes and
their relevance both to the spectroscopic events and to the wind shaping
during the Great Eruption.
I summarize in section ({\ref{sec:summary}}).

{{{
\subsection{Accretion States in $\eta$ Carinae}

Before turning to the new parts, I describe the three different
accretion states of $\eta$ Car that I refer to in this and
the previous papers.
The basic flow structure is as follows.
Both stars, $\eta$ Car (the primary) and its companion, blow winds.
The winds collide, and at one point the momentum flux of
the two winds exactly balance each other, forming a stagnation point.
The stagnation point is located close, but not exactly on,
the line joining the centers of the two stars
(the stagnation point is not exactly on the line
because of the orbital motion).
The shocked material of the primary star cools very fast
(Pittard \& Corcoran 2002; see eq. 1 in Soker 2003), namely,
before the mass moves to a large distance from the stagnation point.
The surrounding pressure then compresses the cooled post-shock
gas to high densities.
Most of the mass is blown by the primary star, hence the
state of the secondary's wind is less of an interest here.

When blowing the wind, in both stars radiation
pressure on the escaping gas overcomes gravitational attraction.
However, this might not be the case with the dense gas near the
stagnation point.
Gravitational force on the dense and slowly moving (much
below escape velocity) gas there might
become large enough to accrete part of the mass back.
Whether accretion occurs at all, and if it does, which of the
two stars accrete most of the gas depends on the state of
$\eta$ Car.
I distinguish between three states.

\begin{enumerate}

\item {\bf The Great Eruption.}
The accretion during the Great Eruption was studied by Soker (2001b).
During the Great Eruption the mass loss rate by the primary was very
high, and the stagnation point was within the companion's (secondary
star) Bondi-Hoyle accretion radius
(see discussion following equation [2] in the next section).
The implication is that along the entire orbit the companion
steadily accreted mass with high specific angular momentum.
Accretion disk was formed, according to the collimated fast wind
(CFW) model, and two jets (or a CFW) were launched.
These jets shaped the two lobes which are now observed
as the Homunculus.
The companion is expected, according to the model, to
accrete $\sim 50 \%$ of the mass blown by the primary,
and blow $\sim 10-20 \%$ of the the accreted gas in the jets.
{{{{ The detection of jets in massive young stellar objects (YSOs)
(Davis et al. 2004) show that jets can be blown by massive stars. }}}}

\item {\bf Apastron passages in present $\eta$ Car.}
In Soker (2003) I proposed that during present apastron
passages the primary itself can accrete a fraction of $\sim 5\%$
of the mass blown over an entire orbit.
This should not be confused with the accretion fraction of
$\sim 50 \%$ by the companion during the 20 years Great Eruption.
The assumption is that part of the cool post-shock primary's wind
material stays bound to the system.
Near apastron passages the primary orbital motion is slow,
and since the the system spends most of the time near apastron,
it was claimed that the primary itself, due to its larger mass,
will accreted back most of the bound mass.
The high specific angular momentum of the accreted gas implies the
formation of an accretion disk, which might blow a CFW.

\item {\bf Periastron passages in present $\eta$ Car.}
This state was not discussed before, and it is a new
idea presented in more detail in the next section.
Basically, the stagnation point of the present wind
of $\eta$ Car is outside the accretion radius of the companion,
and the companion is not expected to accrete mass.
However, if one or two of the following occurs, then accretion is
plausible.
First, if a factor of $\sim 20$ enhanced mass loss rate
near periastron passages occurs (Corcoran et al.\ 2001a),
then the stagnation point is about the same distance from
the companion as the accretion radius (see next section).
Second, when the companion is close to $\eta$ Car, the
wind speed may not reach its terminal speed yet, in
particular if tidal effects of the companion on the
acceleration zone of the wind from $\eta$ Car are important.
The denser $\eta$ Car wind increases the likelihood of accretion.
In any case, the accreted gas is expected to be in the form of
dense blobs (cooled in the post-shock region), leading
to sporadic accretion. This type of flow deserve a 3D
gas-dynamical numerical simulations.

\end{enumerate}

}}}

% ==========================================================
\section{MASS ACCRETION AND WINDS COLLISIONS}
\label{sec:accretion}
% ==========================================================

The main difference between the Great Eruption and the
present flow structure of $\eta$ Car is the
strength of the primary's wind
{{{ (see previous subsection for the different accretion states). }}}
The argument goes as follows (Soker 2001b):
The Bondi-Hoyle accretion radius of the companion, for accretion from
the primary's wind is
\begin{equation}
R_a \simeq {{2 GM_2}\over {v_r^2}} =
0.2
\left({M_2} \over {30 M_\odot} \right)
\left({{v_r} \over {500 \km \s^{-1}}} \right)^{-2}  \AU,
\end{equation}
where $v_r$ is the relative velocity between the wind and the companion.
As discussed in Soker (2001b), $v_r=500 \km \s^{-1}$, is a reasonable value
including both the wind and orbital velocities.
The distance $D_2$ of the stagnation point of the colliding
winds from the companion along the line between the stars, is
given by equating the ram pressures of the two winds $\rho v^2$.
 For spherically symmetric winds
\begin{equation}
D_2= r \beta (1+\beta)^{-1} \simeq r \beta =
a \left( {{1-e^2} \over {1+e \cos \theta}} \right)
\left( {\dot M_2 v_2} \over {\dot M_1 v_1} \right)^{1/2},
\end{equation}
where $r$ is the orbital separation, $\theta$ is the angular
distance along the orbit ($\theta=0$ at periastron), and
$\beta \equiv [(\dot M_2 v_2)/(\dot M_1 v_1)]^{1/2}$.
In the second equality I assumed $\beta \ll 1$.
The stagnation point should be compared with the accretion radius.
For the present winds' parameters of $\eta$ Car, with no
periastron enhanced mass loss rate, taken
from Corcoran et al.\ (2001a),
$v_1=500 \km \s^{-1}$, $\dot M_1=10^{-4} M_\odot \yr^{-1}$,
$v_2=2000 \km \s^{-1}$, and $\dot M_2=10^{-5} M_\odot \yr^{-1}$,
I find $\beta \simeq 0.63$, and $D_2 \sim 0.6 \AU$.
The accretion radius is smaller than $D_2$, hence no continuing
accretion takes place.
During the Great Eruption, when mass loss rate was
much higher, such that at periastron $\beta \simeq 0.02$, and $D_2 < R_a$,
accretion took place near periastron, and indeed along the entire orbit
if the primary wind velocity was lower (Soker 2001b).

If we consider a factor of 20 enhanced mass loss rate lasting 80
days after periastron (Corcoran et al.\ 2001a; but see section \ref{sec:enhanced}),
then $\beta = 0.14 $ and $D_2 \simeq 0.2 \AU$.
The accretion radius is comparable to $D_2$.
Since the shocked material at and near the stagnation point
slows down and cools fast (Soker 2003), it is possible that part
of it will be accreted by the companion during the $\sim 80$~days
of enhanced mass loss rate.
If the enhanced mass loss rate is due to lifting of gas from
the acceleration zone (section {\ref{sec:enhanced}), then the primary
equatorial wind will be very slow.
A slow equatorial flow, $v \lesssim 100 \km \s^{-1}$ is observed indeed
in the vicinity of $\eta$ Car (Zethson et al.\ 1999).
This will not change much the Bondi-Hoyle accretion radius near periastron
because of the high orbital speed, and equation (1) applies.
The distance $D_2$ will increase by a factor of $\sim 2$ due to the lower speed,
and the accretion radius will be somewhat smaller than the stagnation distance
$D_2$.
Overall, I expect that accretion will not occur during most of the periastron
passage, but sporadic accretion of very cold parcels of gas (and dust)
from the wind collision region (see section {\ref{sec:dust}) might occur.

If we consider a slow wind of $v \sim 100 \km \s^{-1}$, then to obscure
X-ray emission during the spectroscopic event, the mass loss rate not need
increase by a factor of $\sim 20$ as required in the X-ray model of
Corcoran et al.\ (2001).
A modest increase of mass loss rate,
by a factor of $\sim 3-5$, is sufficient, and only in and near
the equatorial plane where wind collision occurs.
In that case, the wind interaction is more complicated, as
the orbital speed is higher than the primary's wind speed,
such that it changes the geometry.
The stagnation point is still $D_2 \sim 0.2 \AU$, but its
location is in front of the companion in its orbital
motion, rather than facing the primary.
The relative velocity between the companion and the wind is somewhat
smaller, such that the Bondi-Hoyle accretion radius is somewhat larger.
Sporadic accretion and disk formation may occur in this case as well.

{{{{ As the dense post-shock material flows toward the secondary star
three relevant forces are acting on it: gravity, wind's ram
pressure, and radiation force.
For the parameters of the companion star, wind's ram pressure is not
much smaller, and probably somewhat larger, than radiation pressure
(depending on the unknown secondary luminosity).
It is therefore adequate to consider the force due to wind's ram pressure
and gravity.
I consider a post-shock spherical blob of mass $m_b$, density
$\rho_b$, and radius $r_b$, located at a distance $r$ from the
secondary of mass $M_2$. I also take the blob's density to be $B$
times the wind's density $\rho_b = B \rho_{w2}$.
The magnitude of the gravitational force is given by
\begin{equation}
f_g = \frac{G M_2 m_b}{r^2} = \frac{G M_2}{r^2} \frac {4 \pi}{3}
r_b^3 B \rho_{w2} ,
\end{equation}
where in the second equality I substituted for the blob mass. The
force due to the ram pressure of the wind is given by
\begin{equation}
f_{w2} = \rho_{w2} v^2_{2} \pi r_b^2.
\end{equation}
The ratio of these forces is
\begin{equation}
F \equiv \frac {f_g}{f_{w2}} = \frac{4}{3} \frac{G M_2}{r^2
v^2_{2}}  r_b B \simeq \frac{R_2}{r} \frac {r_b}{r} B ,
\end{equation}
where in the second equality I took the wind speed to be of the
order of the Keplerian speed on the surface of the secondary star,
of radius $R_2$.
For the present system, the first factor on the far right hand
side of the last equation is $R_2/r \gtrsim R_2/D_2 \sim 0.03$,
where the minimum value is obtained for the distance of the
stagnation point $D_2 \sim 0.5 \AU$.
For pressure equilibrium between the ram pressure of the secondary
wind having $v_2=2000 \km \s^{-1}$, and a cool blob at a temperature
of $T_b \sim 10^3-10^4 K$ (see next section) the compression factor
is $B \gtrsim 3 \times 10^4$.
The requirement for the blobs to be accreted, i.e., $F>1$, therefore,
is that its size be $r_b \gtrsim 0.001 r$.
This condition is likely to be met, as the region near the
stagnation point from which mass is accreted is much larger;
many instability modes can be developed with sizes larger
than $0.001r = 0.001D_2$ near the stagnation point.
}}}}

As stated, the sporadic accretion during the spectroscopic event may lead to
disk formation, and jets blown by the companion.
{{{ The dense accreted blobs implies a short period of high
accretion rate.
Therefore, if jets are blown they will be of short duration, i.e.,
high mass loss rate into the jets.
If the jets are narrow, then their }}}
momentum flux
(momentum per unit time per unit area) be larger than the momentum
flux of both the primary's wind and the secondary's wind.
Hence, the jets will expand to large distances
and to the Homunculus.
I would like to speculate that these jets could explain the
high velocity component observed in the X-ray emission of the
iron K-shell line at 6.4~kev (Corcoran et al.\ 2004a).
{{{ It is possible that the origin of the strong
He II 4686 line observed by Steiner \& Demineli (2004) to
reach a velocity of $-400 \km \s^{-1}$ near periastron passages,
is also related to the accretion flow or the outflow. }}}
% ==========================================================
\section{DUST FORMATION IN WINDS COLLISION}
\label{sec:dust}
% ==========================================================

After colliding with the secondary's wind, the wind gas of the
primary star shocks to a temperature of $\sim 3 \times 10^6 \K$
(Corcoran et al.\ 2004b, and references therein).
The cooling time from this temperature to a temperature of
$\sim 10^4 \K$ is much shorter than the flow time
(Pittard \& Corcoran 2002; Soker 2003).
The flow time is defined as
\begin{eqnarray}
\tau_f = {{D_2}\over{v_1}}=3.5
\left( {D_1}\over{1 \AU} \right)
\left( {v_1}\over{500 \km \s^{-1}} \right)^{-1}  ~{\rm days},
\end{eqnarray}
where $v_1$ is the primary's wind speed,
and $D_1$ is the distance of the stagnation point
from the primary star.
However, because the cooling time is much shorter than
the flow time, the post-shock gas is not accelerated much,
and the flow time out of the colliding region is much longer.
For an outflow speed equals the speed of sound at $\sim 10^4 \K$,
the effective flow time is an order of magnitude longer
$\tau_{fe} \sim 1$~month.
For the wind parameters used in the previous section,
{{{ e.g., an enhanced mass loss rate by a factor of
$\sim 20$ during periastron passages (Corcoran et al.
2004), }}} the
hydrogen number density of the pre-shock wind at
distance $D_1$ from the primary is
\begin{equation}
n_{Hw} = 2 \times 10^{10}
\left( \frac{\dot M_1}{10^{-4} M_\odot \yr^{-1}} \right)
\left( \frac{v_1}{500 \km \s^{-1}} \right)^{-1}
\left( \frac{D_1}{1 \AU} \right)^{-2}
\cm^{-1}.
\end{equation}
Behind the shock wave the gas cools and it is compressed by the ram
pressure of the wind to much higher densities.
Equating the thermal pressure of gas at $10^4 \K$ to the ram
pressure of the wind, gives the density in the
cool post-shock gas
$n_p (10^4 \K) \sim 2 \times 10^3 n_{Hw} \sim 4 \times 10^{13} \cm^{-3}$.
Based on figure 11 of Woitke, Kr\"uger \& Sedlmayr (1996),
I find that the cooling time of gas in the density range
$n_H \simeq 10^{11}-10^{14} \cm^{-1}$
from a temperature of $10^4 \K$ to $10^3 \K$ is $\lesssim 1$~month.
This implies that as the post-shock gas flows away from
the stagnation point (in the stage where no accretion to the
companion occurs), it has time to cool to dust-forming
temperatures.
By then, the post-shock gas will be further
compressed to densities of $\sim 10^{14}-10^{15} \cm^{-1}$.
These densities are much higher than in the gas where dust forms
around AGB stars, resulting in more efficient formation of dust,
e.g., large grain.
Therefore, dust with properties different
than the dust around AGB stars might be found in $\eta$ Car environment,
in particular after the system emerge from a spectroscopic event.
{{{ Based on infrared observations (e.g., Whitelock et al. 1994;
Smith et al. 1998), and on extinction (e.g., Hillier et al. 2001)
the presence of large grain in $\eta$ Car was proposed before
(See Smith et al. 2003b for detailed discussion). }}}

Large dust grains are found in the disk around the
binary system HD 44179, with one component being a post-AGB star,
which is located at the center of the
Red Rectangle, a bipolar protoplanetary nebula.
This binary system has an orbital period of $T_{\rm orb} = 322$ days,
a semimajor axis of $a \sin i = 0.32 \AU$, and an eccentricity
of $e=0.34$ (Waelkens {\it et al.} 1996; Waters {\it et al.} 1998;
Men'shchikov et al.\ 2002).
Large grains are also inferred in observations of other
protoplanetary nebulae (e.g., Sahai, Sanchez Contreras, \& Morris 2004).
It is commonly assumed that large dust grains are formed in a long-lived
circumbinary disk around HD 22179 in the Red Rectangle (e.g.,
Jura \& Kahane 1999).
In an earlier paper (Soker 2000b) I suggested that large grains can
be formed in a slow and dense outflow, hence a long lived disk is not
necessary (this view was strengthened by Men'shchikov et al.\ 2002).
Men'shchikov et al.\ (2002) take for the grain-growth time to radius $a_g$
\begin{eqnarray}
t_{\rm gg}  \simeq  0.2
\left( \frac{a_g}{0.2 \cm} \right)
\left( \frac{\rho_g}{2 \g \cm^{-3}} \right)
\left( \frac{n_H}{10^{15} \cm^{-3}} \right)^{-1}
\left( \frac{T}{100 \K} \right)^{-1/2}
\yr.
\end{eqnarray}
This shows that millimeter size grains can indeed formed near
periastron passages of $\eta$ Car.

% ==========================================================
\section{INCREASING ECCENTRICITY}
\label{sec:eccen}
% ==========================================================

 The eccentricity $e$ is reduced by tidal forces on a time scale called
the circularization time and is defined as
$\tau_{\rm circ} \equiv -e/\dot e$.
Therefore, either tidal circularization is weak, or the high eccentricity
of $\eta$ Car, $e = 0.8-0.9$, should be accounted for by a counter effect.
In this section I study these effects in $\eta$ Car.
{{{{ Here I take results from the standard theory of quasi static tides.
However, the tidal interaction in high-eccentricity binary systems
contains contribution from impulsive tidal as well
(Ivanov \& Papaloizou 2004).
In a future work, devoted to tidal interaction, the tidal evolution of the
$\eta$ Car binary system should be calculated with the impulsive tide
contribution included. However, this is beyond
the scope of the present paper. }}}}

In the common tidal model in use, the equilibrium tide mechanism
(Zahn 1977; 1989), the circularization time for evolved stars
is derived by Verbunt \& Phinney (1995).
Here I use the equation as given by Soker (2000a), but change the scaling
to fit $\eta$ Car.
\begin{eqnarray}
\tau_{\rm {circ}} =
2.5 \times 10^6  %4.79 \tim es 10^5
\left( {{f_c} \over {0.2}} \right)^{-1}
\left( {{L} \over {10^7 L_\odot}} \right)^ {-1/3}
\left( {{R} \over {100 R_\odot}} \right)^ {2/3}
\left( {{M_{\rm {env}}} \over {0.01M_1}} \right)^ {-1}
\left( {{M_{\rm {env}}} \over {1 M_\odot}} \right)^ {1/3} \nonumber \\
\times \left( {{M_2} \over {0.25M_1}} \right)^ {-1}
\left( 1+ {{M_2} \over {M_1}} \right)^ {-1}
\left[ {{a(1-e)} \over {3.6R}} \right]^ {8}
\yr ,
\label{eq:taucir}
\end{eqnarray}
where $L$, $R$ and $M_1$ are the luminosity, radius, and total mass of the
primary star, $M_{\rm {env}}$ is the primary's envelope mass,
and $f_c \simeq 1$ is a dimensionless function of the eccentricity.
The envelope is assumed to be convective. If it is radiative,
then the circularization time is longer.
For a slow rotating star the function $f_c(e)$
is given by (Hut 1982)
\begin{eqnarray}
f_c(e) \simeq (1-e)^{3/2}
\left(1+\frac{15}{4}e^2 + \frac{15}{8}e^4 + \frac{5}{64}e^6 \right),
\label{eq:fce1}
\end{eqnarray}
where I moved a factor of $(1-e)^8$ from the usual definition
of $f(e)$ to the last term producing a more transparent equation,
and when $e=0.9$, $f_c=0.17$.
This shows that tidal interaction will not reduce much the eccentricity, even
for a more massive envelope, and even if during part of
the time the primary's radius is larger than its present value.

On the other hand, the eccentricity may increase because of
the interaction, via enhanced mass loss rate at periastron passages
or via a circumbinary disk.
The change in eccentricity due to an isotropic mass loss
(the derivation is applicable for an axisymmetric mass loss as well)
is given by (Eggleton 2005)
\begin{eqnarray}
\delta e =
{{\vert \delta M \vert} \over {M}} (e + \cos \theta),
\label{eq:ecc1}
\end{eqnarray}
where $\delta M$ is the mass lost from the binary in the stellar wind
at the orbital phase $\theta$ (hence $\delta M < 0$), $M$ is the total
mass of the binary system, and $\theta$ is the orbital angle, measured
from periastron, of the position vector from the center of mass to
the secondary.
The derivation of equation (\ref{eq:ecc1}) assumes that
$\delta M(\theta)=\delta M (-\theta)$.
As in Soker (2000a) I assume that in addition to its constant mass
loss rate over the orbital motion $\dot M_w$, the primary star
loses an extra mass $\delta M_p$ in a short time during the periastron
passage, $\cos \theta =1$.
 The total mass being lost in one orbital period
$T_{\rm orb}$, is $\Delta M_o = \dot M_w T_{\rm orb} + \delta M_p$.
I define the fraction of the mass being lost at periastron
\begin{eqnarray}
\beta \equiv {{\delta M_p}\over{\Delta M_o}}.
\label{eq:ecc2}
\end{eqnarray}
Under the assumption that the fraction of mass lost at
periastron passage $\beta$ does not change during the
evolution, the relation between initial and final eccentricity is
(Soker 2000a)
\begin{eqnarray}
{{1+e}\over{1+e_i}} =
\left( {{M_i}\over{M}} \right)^{\beta},
\label{eq:ecc3}
\end{eqnarray}
where $e_i$ and $M_i$ are the initial eccentricity and total binary
mass, respectively, and $M$ is the final total mass.
For example, if on average during the {{{ entire }}}
evolution {{{ (mainly before the Great Eruption), }}}
half of the mass was
lost during periastron passages, i.e., $\beta=0.5$,
and the total system mass changed from $M_i=180 M_\odot$ to a final
(present) mass of $M=140 M_\odot$, then
$(1+e)/(1+e_i)= 1.13$.
For example, when $e_i=0.7$, $e=0.9$.
This shows that an enhanced periastron mass loss rate, if occurring,
can increase somewhat the eccentricity.
In any case, equations (\ref{eq:taucir}) and (\ref{eq:ecc3}) explain
why the system eccentricity was not reduced substantially by tidal forces.

Mass transfer can also change eccentricity.
The change in eccentricity due to a mass  $\delta M_{\rm tran}$
transferred from the primary to the secondary is given by (Eggleton 2005)
\begin{eqnarray}
\delta e = 2 \delta M_{\rm tran}
\left( {{1}\over{M_1}} - {{1}\over{M_2}} \right) (e+\cos \theta).
\label{eq:ecc4}
\end{eqnarray}
Since enhanced mass transfer may occur during periastron passage,
we see that the eccentricity will decrease if $M_1 > M_2$, as is the
case in $\eta$ Car.
Since for $\eta$ Car $M_1 \gg M_2$, we find from equations (\ref{eq:ecc1})
and (\ref{eq:ecc4}) that in order for the eccentricity not to decrease,
the secondary must not accrete more than a fraction of $\sim M_2/2M_1$
of the mass lost near periastron passages.
During the Great Eruption, this situation may have been different
because of the dense and somewhat slower wind (Soker 2001b), and the
companion accreted mass along its {\it entire} orbit.
This tends to reduce the orbital separation, but does not change
much the eccentricity.

As I mentioned in earlier papers (Soker 2003), $\eta$ Car is not unique.
The most relevant system regarding this section is the binary system
HD 44179 (period of $T_{\rm orb} = 322$ days,
a semimajor axis of $a \sin i = 0.32 \AU$,
and an eccentricity of $e=0.34$; Men'shchikov et al.\ 2002),
which is located at the center of the
Red Rectangle, a bipolar protoplanetary nebula
(Cohen et al.\ 2004).
In Soker (2000a) I argue that enhanced mass loss during periastron passages
can account for the non-zero eccentricity of HD 44179.
Waelkens et al.\ (1996), Waters et al.\ (1998), and
Men'shchikov et al.\ (2002) attribute the non-zero eccentricity to
the presence of a disk around the binary system.
In this ``external disk'' mechanism, tidal interaction between
the binary system and the circumbinary disk enhances the eccentricity,
as is the model to explain eccentricities in young stellar binaries
(Artymowicz et al.\ 1991; Artymowicz \& Lubow 1994).

Whether the mechanism for enhancing the eccentricity is an enhanced
mass loss rate during periastron passages, or the external disk
mechanism, the mechanism appears to be the same in $\eta$ Car
and the Red Rectangle.
The periastron mechanism implies that enhanced mass loss rate
at periastron passage was a strong effect.
The external disk mechanism implies that a relatively massive
disk {{{ might be }}} present, or {{{ might have been }}} present
before the Great Eruption, close to the binary system of $\eta$ Car.

The time required for the secondary to spiral in to the
primary is $\tau_{\rm in}= \tau_{\rm circ} /16$.
This is a long time in $\eta$ Car, during which the orbital
separation increases because of mass loss.
Hence, I do not expect the secondary to enter the primary envelope,
unless the primary's envelope substantially swells.

% ==========================================================
\section{TIDAL SPIN UP}
\label{sec:tidal}
% ==========================================================

 The synchronization time between orbital angular velocity and
primary's spin angular velocity is
$\tau_{\rm {syn}} \simeq  (1+M_2/M_1)(M_2/M_1)^{-1}(I/M_1R^2)
(R/a)^2 \tau_{\rm {cir}} $, where $I$ is the primary's moment of inertia.
For a crude estimate, I approximate the envelope density profile of
$\eta$ Car by $\rho \propto r^{-2}$, as for stars on the upper AGB,
where $r$ is the radial distance from the star's center.
For that density profile $I=(2/9) M_{\rm env} R^2$.
The scaled expression for the synchronization time reads
\begin{eqnarray}
\tau_{\rm {syn}} \simeq 2 \times 10^ 3
\left( {{f_s} \over {0.2}} \right)^{-1}
\left( {{L} \over {10^7 L_\odot}} \right)^ {-1/3}
\left( {{R} \over {100 R_\odot}} \right)^ {2/3}
\left( {{M_{\rm {env}}} \over {1 M_\odot}} \right)^ {1/3}
% \nonumber \\
\left( {{M_2} \over {0.25 M_1}} \right)^ {-2}
\left[ {{a(1-e)} \over {3.6R}} \right]^ {6}
\yr.
\label{eq:spin1}
\end{eqnarray}
For a slowly spinning star the function $f_s(e)$ reads (Hut 1982)
\begin{equation}
f_s(e) \simeq (1+e)^{-6}
\left(1+\frac{15}{2}e^2 + \frac{45}{8}e^4 + \frac{5}{16}e^6 \right);
%\label{eq:syn2}
\end{equation}
when $e=0.9$, $f_s(0.9)=0.23$.
The short synchronization time implies that even without mass loss,
which further slows down the envelope (Soker 2004a),
the primary star will reach synchronization with the orbital motion
during secular evolution.
Namely, the primary will spin with a period of $\sim 5 \yr$, which is
$\ll 10\%$ of its break-up angular speed.
{{{{ The discussion above is based on the standard quasi-static
tidal theory.
However, in the impulsive tidal interaction theory, the companion can
spin-up the primary to a rotation speed of $> 0.1$ times its break-up
speed (Ivanov \& Papaloizou 2004).
In any case, the mass loss from the primary star in the Great Eruption
will substantially slow down the primary rotation, and the mass loss
geometry is not expected to deviate much from spherical one
(Soker 2004a). }}}}
It is the binary companion which is behind the slow, dense
equatorial mass flow, and it is the companion that
ejected the collimated fast wind (jets) which inflated the
two lobes during the Great Eruption (Soker 2001b).

% ==========================================================
\section{ENHANCING MASS LOSS RATE}
\label{sec:enhanced}
% ==========================================================

A suggestion that enhanced mass loss rate near periastron passages
accounts for the spectroscopic event (Davidson 1999;
Ishibashi et al.\ 1999; Corcoran et al.\ 2001) appears fundamental
to some models as well as influencing some key issues in this paper.
It is discussed in this section.

What is the mechanism by which the companion increases, if it does,
the mass loss rate near periastron passages?
Some works related to other astrophysical objects propose enhanced
mass loss rate as a result of the presence of a companion.
The effect is basically that the companion spins up the primary
or rises a tidal bulge.
A commonly used expression for the enhanced mass loss rate in binary
systems is
\begin{eqnarray}
\dot M_1 = A_1 \frac {R_1 L_1}{M_1}
\left[1 + B_L \left( \frac {R_1}{R_L} \right)^\gamma \right],
\label{eq:dotme2}
\end{eqnarray}
where $M_1$, $R_1$, and $L_1$, are the mass-losing giant stellar mass,
radius, and luminosity, respectively,
$R_L$ is the radius of a sphere that has the same volume as
the Roche lobe of the primary mass losing star,
and $A_1$, $B_L$ and $\gamma$ are constants.
Different values for $B_L$ and $\gamma$ are quoted in the literature:
Tout \& Eggleton (1988) use $B_L=10^4$, $\gamma=6$, and if
$R_g/R_L>0.5$ then they set 0.5 for this ratio.
Han et al.\ (1995) prefer $B_L \simeq 500$ and $\gamma=6$,
while Han (1998) argue for $B_L \simeq 1000$ and $\gamma=6$.
Frankowski \& Tylenda (2001) argue for a more complicated
expression, which basically has a very low value of $B_L$ and $\gamma=3$;
however, their numerical calculations yield a much faster increase
in the mass loss rate as the giant is close to filling its Roche lobe.

The derivation of equation (\ref{eq:dotme2}), which uses the Roche
lobe radius, assumes that the mass losing star corotates with
the orbital motion.
This cannot be the case in a highly eccentric
orbit such as in the case of $\eta$ Car, where the
orbital angular speed is much larger than the primary spin's
angular velocity near periastron passages.
Instead I take the point where the gravity forces of the two
stars are equal.
For an eccentricity of $e=0.9$ and semimajor axis of $a=16.6 \AU$,
at periastron passage $r=360 R_\odot$, and the gravity of the two stars
are equal at a distance from the primary's center of
$r_1 \sim 240 R_\odot$.
The radius of $\eta$ Car is $R_1 \sim 100 R_\odot$,
such that $R_1/r_1 \sim 0.4$.
Replacing $R_1/R_L$ by $R_1/r_1$ in equation (\ref{eq:dotme2}),
gives the enhanced mass loss rate values of $\sim 40$,
$3$, and $5$,
for the parameters of Tout \& Eggleton (1988),
Han et al.\ (1995), and  Han (1998), respectively.
The prescription of Frankowski \& Tylenda (2001) gives negligible
enhanced mass loss rate.

Frankowski \& Tylenda (2001) try to find a mechanism to enhance
the mass loss rate and find that it is not clear how the mechanism
can substantially enhance the mass loss rate for $R_L \gtrsim 2 R_1$.
One possible way out of this problem is by using a mechanism that does
not affect the stellar surface as much as it affects the region
from which the wind is accelerated.
This region in giants can be quite large, in particular if
dust formation occurs.
Shocks from stellar pulsation can enhance or trigger dust formation
(Woitke, Goeres, \& Sedlmayr 1996).
It might be, therefore, that the role of the companion in the case of
$\eta$ Car is to make dust formation favorable above the atmosphere
of $\eta$ Car, by enhancing gas density or inducing shocks.
In the case of enhancing density, the relevant radius
for the tidal interaction is not $R_{\rm tid}=R_1$, but rather the
radius where dust might form, which is $R_{\rm tid} \sim 1.5-2.5 R_1$.
Since the height of a tidal bulge is proportional to $(R_{\rm tid}/r)^3$,
where $r$ is the orbital separation, an increase by a factor of
$\sim 2$ in $R_{\rm tid}$ can make a huge difference.
Is it possible that dust can form as close to the surface of a hot
star as the primary star of $\eta$ Car is ($\sim 20,000 \K$)?
I note that dust formation occurs near the surface of hot
R Coronae Borealis (RCB) stars.
RCB are rare hydrogen-deficient carbon-rich supergiants which
undergo very spectacular declines in brightness of up to
8 mag at irregular intervals as dust forms along the line of sight
(Clayton 1996).
The hot RCB stars have an effective temperature of $\sim 18,000 \K$,
a luminosity of $L \sim 10^4 L_\odot$, hence a radius of
$R \simeq 10 R_\odot$.
Their surface temperature and gravity are similar to that of
$\eta$ Car, but they are carbon rich, facilitating a lot dust
formation.
In any case, hot RCB stars show that dust can be formed close
to the surface of hot stars.

There is another difficulty.
Even if the wind is slow, $v_1 \sim 100 \km \s^{-1}$,
the wind momentum flux to radiation momentum flux ratio
for an enhanced mass loss rate of
$2 \times 10^{-3} M_\odot \yr^{-1}$, and luminosity of
$L_1 = 3 \times 10^6 L_\odot$,
is quite high, $\zeta \equiv \dot M_1 v_1 / (L_1/c) \simeq 3$.
That is, the momentum flux in the wind is quite high, and
it is not easy to explain this high momentum by radiation pressure.
As stated in Section \ref{sec:accretion}, the enhanced mass loss
rate for a slow wind might be larger by only a factor
of $\sim 4$, and near the equatorial plane.
If the mass loss is spherical, then $\zeta < 1$.
In these circumstances, the supposition is that the enhanced
mass loss rate occurs near the equatorial plane, where
less radiation is available for mass ejection.

Whitelock et al.\ (2004) argue that dust absorption has negligible
role in the spectroscopic events, hence dust cannot
absorb and scatter a large fraction of the radiation emitted
by the primary star.

Considering the problems mentioned above, it is quite possible that
the spectroscopic event is mainly a coverage of the region around the companion,
as proposed by Whitelock et al.\ (2004).
As discussed in Section \ref{sec:accretion},
large quantities of dust could be formed in the region near the companion,
possibly engulfing the companion as it deflects the flow around itself,
or even accreting mass.
Most of the radiation from the primary star is not blocked,
but some fraction of the radiation is blocked.
To estimate the magnitude of these effects, in the next section
the ionization of the gas by the companion is discussed.

This section is summarized by emphasizing again that it is not
clear whether substantial increase in the mass loss rate from
the primary occurs near periastron passages, or whether the secondary
enters a more or less static dense medium.
If the mass loss does increase, there is a larger question regarding
the mechanism behind the increased mass loss rate.
This question is an important open question regarding
other astrophysical objects, for example, the Red Rectangle.

% ==========================================================
\section{IONIZATION BY THE COMPANION}
\label{sec:ionization}
% ==========================================================

The ionization of gas in the nebula of $\eta$ Car was discussed in
a previous paper (Soker 2001a).
In that paper I propose ionization shadows as an explanation for
the formation of the long and narrow strings of $\eta$ Carinae
(see Weis, Duschl \& Chu 1999), and I show that the companion
could play a significant role in the ionization of the strings'
surroundings.
The main ionization was attributed to the time the companion is away
from periastron, which is most of the orbital period.
Here I examine the situation near periastron.

The flow structure is very complicated near periastron passages,
and 3D numerical simulations are required to find the shape of the
ionized region.
I therefore use a simple flow structure (as in Soker 2001a),
wherein the primary's wind flows undisturbed around the companion,
and I consider only the ionization of the primary wind,
neglecting the companion wind.
Although ideal, the derived expression gives a good indication
of the main effect.
Because of the complicated flow structure near the equatorial plane,
I examine the ionization in the polar direction from the companion.
Let $r$ be the distance between the companion and the primary, and
$h$ the distance above the poles of the companion, i.e.,
perpendicular to the equatorial plane.
The density of the primary's wind above the companion is
$\rho(h)= \dot M_1 / [4 \pi (r^2 + h^2) v_1]$.
The total recombination rate, assuming spherical geometry
for the sake of the calculation, along the direction $h$ is given by
\begin{eqnarray}
\dot N = \int_{h_0}^{h_i} 4 \pi \alpha n_e(h) n_i(h) h^2 dh.
\label{eq:recomb1}
\end{eqnarray}
where $n_i$ and $n_e$ is the ion and electron density, respectively,
and $\alpha$ the recombination coefficient.
Performing the integration from inner distance $h_0$ to
the edge of the ionization front $h_i$ gives
\begin{eqnarray}
\dot N = 4 \pi \alpha
\frac{n_e}{\rho} \frac{n_i}{\rho}
\left( \frac{\dot M_1}{4 \pi v_1} \right)^2
\frac{1}{2r} \left[H_0(1+H_0^2)^{-1}-H_i(1+H_i^2)^{-1}
+\tan^{-1} H_i - \tan^{-1}H_0 \right],
\label{eq:recomb2}
\end{eqnarray}
where $H\equiv h/r$.

To find the distance of the ionization front in units of the
orbital separation $H_i$, I assume that
$H_0 \ll H_i \ll 1$, which is correct near periastron when
the companion enters a dense medium (see below).
Equating $\dot N$ to the ionizing photon luminosity of the companion
$\dot S_2$, scaled as in Soker (2001a), and using the approximation
above, yields, after some manipulation
\begin{eqnarray}
H_i \equiv \frac{h_i}{r} \sim 0.2
\left( \frac{r}{2 \AU} \right)^{1/3}
\left( \frac{v_1}{100 \km \s^{-1}} \right)^{2/3}
\left( \frac{\dot M_1}{2 \times 10^{-4} M_\odot \yr^{-1}} \right)^{-2/3}
\left( \frac{\dot S_2}{2 \times 10^{48} \s^{-1}} \right)^{1/3}.
\label{eq:recomb3}
\end{eqnarray}
The numerical value in the last equation will not change
if we take $v_1=500 \km \s^{-1}$ (no slower wind near
periastron), and at the same time increase the mass loss rate
to $\dot M_1 = 10^{-3} M_\odot \yr^{-1}$.

Although the last equation was derived under strong simplifications
and assumptions, it does provide an insight into the ionization zone.
As long as the companion is away from periastron, and the wind
parameters are the present ones, i.e., mass
loss rate is $\dot M_1 \sim 10^{-4} M_\odot \yr^{-1}$ and
$v_1 \sim 500 \km \s^{-1}$, the ionization zone is large,
$h_i \gtrsim r$, and an accurate solution of equation
(\ref{eq:recomb2}) is required.
If, however, near periastron the companion enters into a
much higher density zone, caused by enhanced mass loss rate
or slower velocity, then the ionization zone shrinks to
a size $h_i \ll r$.
This implies that the ionization zone could be hidden by
dense primary's wind, in particular by the dense gas
in the region where the two winds interact.

Finally, I note that a flow structure where a companion ionizes the
primary's wind is not unique to $\eta$ Car.
For example, it occurs in symbiotic binary systems and some
protoplanetary nebulae.
In the protoplanetary nebula M2-9 there is a positional shift of
bright knots in the inner nebular lobes on a period of 120 years
(Doyle et al.\ 2000).
This side to side departure from axisymmetry is explained in terms of a
revolving ionizing source, i.e., an ionization by one star in
a binary system (Livio \& Soker 2001).
Livio \& Soker (2001) show that the interaction between the slow,
AGB star's wind, and a collimated fast wind from the white dwarf companion
clear a path for the ionizing radiation in one direction,
while the radiation is attenuated in others.
Similar flow and radiation structure (but not exactly identical in the
geometry and winds parameters) was qualitatively applied to the side
to side variation in the UV images before and after the spectroscopic event
in $\eta$ Car (Smith et al.\ 2004a).
In $\eta$ Car the UV images are of the near UV, with energy below the hydrogen
ionization threshold.
This UV radiation can expand to larger distances than the
ionizing radiation studied in this section.
{{{ It is interesting to note that the general bipolar structure of
M2-9 was compared to the bipolar structure of $\eta$ Car by
Smith (2003), and the nuclear spectrum of M2-9 was compared to that
of $\eta$ Car by Balick (1989) and Allen \& Swings (1972). }}}

% ==========================================================
\section{SUMMARY}
\label{sec:summary}
% ==========================================================

The main goal of this paper is to further explore the role
of the binary companion in $\eta$ Car, and by that to strengthen
the link between $\eta$ Car and other binary systems losing
mass at high rates and having bipolar nebulae around them.
{{{{ This, I hope, will motivate more 3D gasdynamical numerical
simulations of the interaction of the two stars and their winds. }}}}

A unified model for shaping bipolar nebulae (nebulae composed of
two lobes with an equatorial waist between them) around mass losing
stars has been emerge in recent years (Soker 2004b).
In this model (Morris 1987, 1990), the two opposite lobes are
shaped by a binary companion accreting mass from the mass-losing
primary, and then blowing two jets.
The two jets then inflate the bubbles, as in many other
astrophysical objects, such as clusters of galaxies
(Soker 2004b and references therein).
Note that I refer here only to nebulae having two large lobes.
Most elliptical planetary nebulae which have no
large lobes, for example, were not shaped by this process.
A second goal was to compare the present effects of certain
processes with the effects of these processes during the
Great Eruption of a century and a half ago.

The main finding of this paper can be summarized as follows.

\begin{enumerate}

\item As far as the binary interaction goes, the main question
regarding the spectroscopic event is on the effect of the
companion on the mass loss process near periastron passages.
Does the companion increase the mass loss rate from the primary?
If it does, by how much? How does it change the geometry of the
mass loss process, i.e., does the wind becomes slower near the
equatorial plane?
As discussed in Section (\ref{sec:enhanced}), periastron enhanced
mass loss rate was discussed in relation to other astrophysical
objects, e.g., in the central binary system of the bipolar
protoplanetary nebulae the Red Rectangle.
However, if mass loss rate is indeed enhanced, a mechanism is
yet to be identified for that to occur.
A promising mechanism could be one in which the companion
tidaly facilitates dust formation (Section \ref{sec:enhanced}).

\item One of the results of periastron enhanced mass loss rate
is a departure from axisymmetry in the equatorial plane.
In the first type of departure from axisymmetry
the mass loss rate near periastron and apastron
causes a permanent differences between the two sides of the nebula.
This is observed in $\eta$ Car {{{ (e.g., Smith 2002; }}}
see discussion in Soker 2001b), and in the inner
region of the Red Rectangle (e.g., Tuthill et al.\ 2002;
Miyata et al. 2004), which is known to harbor
a binary stellar system at its center.
If the major axis of the eccentric orbit is perpendicular to our
line of sight (Smith et al.\ 2004a), then the observed departure
in $\eta$ Car is small.

\item There is another type of departure from axisymmetry, where
there is a side to side variation as the two stars
orbit the center of mass.
This is observed in $\eta$ Car in the UV (Smith et al.\ 2004a) in a
manner resembling that of the protoplanetary nebulae M2-9, where the
orbital period is $120$~years (Doyle et al.\ 2000).
The explanation for this side to side variation during the orbital motion
is the opacity of the dense wind blown by the mass-losing star, which
allows the radiation from the companion to propagate only in the
other direction. This radiation comes from a hot white dwarf in
M2-9 (Livio \& Soker 2001), and from an O star in $\eta$ Car
(e.g., Smith et al.\ 2004a).

\item The winds from the two stars collide, and the post-shock
primary's wind cools on a time scale much shorter than the flow time.
Even without enhanced or slow wind, near periastron passages
a dense, cold region is formed.
Large dust grains, of the size of a millimeter size,
could be formed there (Section \ref{sec:dust}).
Large dust grains are observed in the Red Rectangle;
the popular dust formation site there is a long-lived
circumbinary disk, although a slow, dense flow as the site
for large grain formation was suggested in the Red Rectangle as
well.

\item If mass loss rate is indeed enhanced near periastron passages,
or the wind is slower,
the companion could sporadically accrete mass from the dense,
cold region in the interaction region of the two winds
(Section \ref{sec:accretion}).
Such accretion events may lead to the formation of an accretion
disk around the companion, which could then blow two jets.
The jets might expand to large distances, revealing themselves as
fast velocity components.
In Section (\ref{sec:accretion}) I speculate that these jets
may be connected to the high velocity component observed in
the X-ray emission of the iron K-shell line at 6.4~kev
(Corcoran et al.\ 2004a).
I note that during the Great Eruption, the companion accreted mass
along its entire orbit and probably blew the jets
that shaped the Homunculus (Soker 2001b; see section 1.2 here).

\item
Periastron enhanced mass loss rate increases eccentricity.
If, during the evolution, an average mass fraction of $\beta$ was
lost near periastron passages, then the eccentricity
was increased according to equation (\ref{eq:ecc3}).
In any case, tidal interaction did not much decrease the
eccentricity (Section \ref{sec:eccen}), and if the eccentricity
was large initially, it remained so.
High eccentricity, although not as high as in $\eta$ Car,
is found also in the bipolar nebula the Red Rectangle.

\item Because of short synchronization time (equation \ref{eq:spin1}),
the the primary spin period is $\lesssim 5 \yr$, meaning that its
rotation speed is much below its break-up rotation.
Hence, effects of its rotation on the axisymmetrical mass loss
geometry are expected to be small (Soker 2004a)

\item During most of the orbit the companion is likely to
ionize a large volume of the primary's wind (Soker 2001a).
This process of ionization by a companion occurs
also in symbiotic binary systems.
{{{ Some similarities of the nuclear emission of $\eta$ Car
with symbiotic systems and planetary nebulae was noted before
(e.g., Balick 1989; Smith 2003). }}}
However, near periastron passages, if the $\eta$ Car primary's
wind is slow and/or mass loss rate is higher, the ionized
region shrinks, such that it can be eclipsed,
partially or fully, by the primary's dense wind
(Section \ref{sec:ionization}), as the X-ray emission is
eclipsed (Corcoran et al.\ 2004b).

\end{enumerate}

\bigskip

% {{{ I thank an anonymous referee for useful comments.  }}}
This research was supported in part by a grant from the Israel
Science Foundation.

\end{document}